\documentclass[aps,prd,superscriptaddress,longbibliography,amssymb,twocolumn]{revtex4-1}
\usepackage[utf8]{inputenc}
\usepackage[pdftex]{graphicx}
\usepackage{amsmath}
\usepackage[amssymb,Gray]{SIunits}
\usepackage{physics}
\usepackage{hyperref}
\hypersetup{breaklinks=true}
\newcommand{\rmi}{\ensuremath{\mathrm{i}}}
\newcommand{\rme}{\ensuremath{\mathrm{e}}}
\newcommand{\rmd}{\ensuremath{\mathrm{d}}}

\newcommand{\nnl}{\nonumber\\}

\begin{document}

\title{Self-gravitational dephasing of quasi-classical Stern-Gerlach trajectories}


%
\author{Andr\'e Gro{\ss}ardt}
\email[]{andre.grossardt@uni-jena.de}
\affiliation{Institute for Theoretical Physics, Friedrich Schiller University Jena, Fr\"obelstieg 1, 07743 Jena, Germany}


\date{\today}

\begin{abstract}
The nonlinear Schr\"odinger--Newton equation, a prospective semiclassical alternative to a quantized theory of gravity, predicts a gravitational self-force between the two trajectories corresponding to the two $z$-spin eigenvalues for a particle in a Stern-Gerlach interferometer. To leading order, this force results in a relative phase between the trajectories. For the experimentally relevant case of a spherical particle with localized wave function, we present a re-derivation of that phase which is both rigorous in its approximations and concise, allowing for simple but accurate experimental predictions.
\end{abstract}


\maketitle



\section{Introduction}

In light of the inconclusive theoretical evidence~\cite{mattinglyQuantumGravityNecessary2005,kieferQuantumGravity2007} for the need to quantize the gravitational interaction, there has been some interest in an experimental evaluation whether gravity is quantized. One of the most commonly discussed alternative models is semiclassical gravity~\cite{grossardtThreeLittleParadoxes2022,giuliniCouplingQuantumMatter2023} or its nonrelativistic limit, the nonlinear Schr\"odinger-Newton (SN) equation~\cite{diosiGravitationQuantummechanicalLocalization1984,penroseQuantumComputationEntanglement1998,carlipQuantumGravityNecessary2008, giuliniGravitationallyInducedInhibitions2011,bahramiSchrodingerNewtonEquationIts2014}. It predicts that a point particle with spatial wave function $\psi$ is subject to a gravitational potential $U_\psi$ sourced by the mass distribution $m \abs{\psi}^2$; for the center of mass of a composite object the gravitational potential is a double convolution of $\abs{\psi}^2$ with the internal mass distribution $\rho$~\cite{grossardtEffectsNewtonianGravitational2016}. Experimental tests have been proposed, for instance, in optomechanical systems~\cite{yangMacroscopicQuantumMechanics2013,grossardtOptomechanicalTestSchrodingerNewton2016}.

More recently, ideas to witness whether gravity is capable of inducing entanglement between remote systems~\cite{kafriClassicalChannelModel2014,boseSpinEntanglementWitness2017,marlettoGravitationallyInducedEntanglement2017} have gained some traction, with an ongoing debate about the conclusiveness of a potential positive outcome~\cite{donerGravitationalEntanglementEvidence2022, fragkosInferenceQuantizationGravitationally2022, palExperimentalLocalisationQuantum2021, reginattoEntanglingQuantumFields2019}. While these entanglement tests generally require two or more adjacent interferometers, it has been pointed out~\cite{hatifiRevealingSelfgravitySternGerlach2020,grossardtDephasingInhibitionSpin2021} that nontrivial effects of the SN equation occur already in a single Stern-Gerlach interferometer due to the gravitational self-force between the trajectories. These works study in great detail (as witnessed by their length of 37 pages each) the self-gravitational effects, also in the theoretically more challenging and, therefore, more interesting regimes where the wave function spread compares to or exceeds the size of the particle.

A recent article~\cite{aguiarProbingGravitationalSelfdecoherence2023} provides a much more compact derivation of only the gravitational phase shift in the experimentally relevant situation of a particle in superposition of two trajectories with sharply peaked spatial distributions. Unfortunately, it explicitly neglects the self-energy contribution which can in principle both increase or decrease the total phase difference, but for realistic parameters will dominate the predicted phase.

Here, we re-consider the same experimental situation of a homogeneous, sphere-shaped spin-$\tfrac12$ particle of radius $R$ with a split in two spin-dependent trajectories at distance $d > R$ and described by wave packets with a variance much smaller than $R^2$. Employing the rigorous approximation methods developed in ref.~\cite{grossardtDephasingInhibitionSpin2021}, however, restricting calculations to the appropriate limit, enables us to present a concise derivation with a simple and intuitive final result for the predicted dephasing.

In section~\ref{sec:sep} we begin with setting up the model, followed by a transformation that gets rid of the accelerating potential in section~\ref{sec:comoving}. We then solve the SN equation for Gaussian initial conditions and derive an approximation for the phase shift in section~\ref{sec:gauss}, concluding with a discussion of the results in section~\ref{sec:discussion}.

\section{Separating spin and position}\label{sec:sep}

We consider a spin-$\tfrac12$ particle in a Stern-Gerlach interferometer, i.e., subject to a spin-dependent force. We express the normalized state vector in the $z$-spin basis as
\begin{equation}\label{eqn:state}
 \ket{\Psi}_t = \cos \alpha_t \ket{\uparrow}\ket{\psi_\uparrow}_t + \sin \alpha_t \ket{\downarrow}\ket{\psi_\downarrow}_t
\end{equation}
with $\alpha_t \in \mathbb{R}$. All phases can be absorbed into the spatial wave functions $\ket{\psi_{\uparrow\downarrow}}_t$ (taken to be initially equal). Assuming a $z$-spin proportional coupling to a homogeneous, time-dependent force, the state vector~\eqref{eqn:state} solves the Schr\"odinger equation
\begin{equation}\label{eqn:schroedinger}
 \rmi \hbar \partial_t \ket{\Psi}_t
 = -\frac{\hbar^2}{2m} \partial_z^2  \ket{\Psi}_t
 + \left( U_\Psi - \sigma_z F_t z \right) \ket{\Psi}_t \,,
\end{equation}
where the time dependent homogeneous force $F_t$ determines the spin-dependent trajectories and $U_\Psi$ is the state-dependent (thus nonlinear) gravitational potential. Equation~\eqref{eqn:schroedinger} conserves the $z$-spin, hence $\alpha_t \equiv \alpha$ remains constant.

We linearize this equation by first solving for $\ket{\Psi_0}_t$ in the absence of $U_\Psi$ and then solving equation~\eqref{eqn:schroedinger} for $U_{\Psi_0}$ calculated from the gravity-free solution. This is a good approximation as long as the spatial wave functions $\ket{\psi_{\uparrow\downarrow}}$ in $\ket{\Psi_0}_t$ and $\ket{\Psi}_t$ do not significantly differ in more than a phase. Note that this allows for arbitrary relative phases between the spin-up and spin-down parts. The resulting linearized equation separates:
\begin{equation}\label{eqn:spatial-schroedinger}
 \rmi \hbar \partial_t \ket{\psi_{\uparrow\downarrow}}
 = -\frac{\hbar^2}{2m} \partial_z^2 \ket{\psi_{\uparrow\downarrow}}
 + V_{\uparrow\downarrow} \ket{\psi_{\uparrow\downarrow}} \,,
\end{equation}
where $V_{\uparrow\downarrow} = U \mp F_t z$, with the gravitational potential $U$ calculated from the wave functions $\psi^{(0)}_{\uparrow\downarrow}$ that solve equation~\eqref{eqn:spatial-schroedinger} for $U=0$. This can in principle be iterated to higher orders by re-inserting the solution to calculate a corrected potential, but we only consider the first order effects, which amounts to neglecting terms of $\order{G^2}$, $G$ being the gravitational constant.

Despite resulting in a linear Schr\"odinger equation, the gravitational potential $U$ generally depends on the shape of the wave function as well as the mass distribution of the particle in a complicated way~\cite{grossardtDephasingInhibitionSpin2021}. For wave functions $\psi_{\uparrow\downarrow}$ sharply peaked (compared to the particle radius) around $u^{\uparrow\downarrow}$, and for the particle being modelled as a homogeneous sphere of radius $R$, the gravitational potential becomes
\begin{subequations}\label{eqn:potential}\begin{equation}
 U(u^\uparrow,u^\downarrow;z) = -\frac{G m^2}{R} \sum_{\sigma \in \{\uparrow,\downarrow\}} c_\sigma \, \Xi\left(\frac{\abs{z - u^\sigma}}{2R}\right)
\end{equation}
where we write $c_\uparrow = \cos^2(\alpha)$ and $c_\downarrow = \sin^2(\alpha)$ for convenience and introduce the function
\begin{equation}
 \Xi(\xi) = \begin{cases}
\frac{6}{5} - 2 \xi^2 + \frac{3}{2} \xi^3 - \frac{1}{5} \xi^5 &\quad \text{for } \, \xi \leq 1 \\
\frac{1}{2\xi} &\quad \text{for } \, \xi \geq 1 \,.
\end{cases}
\end{equation}\end{subequations}
The potential~\eqref{eqn:potential} has a simple interpretation: it is the gravitational potential energy of a test particle of mass $m$ in a Newtonian potential which is a superposition of two potentials belonging to spheres of radius $R$ with masses $m\cos^2(\alpha)$ and $m\sin^2(\alpha)$, located at positions $u^\uparrow$ and $u^\downarrow$, respectively.

\section{Co-moving frame}\label{sec:comoving}

In order to deal with the time dependent force $F_t$ acting on each spin part with opposite signs, we perform a Galilean coordinate transformation into the co-moving frame with each spin trajectory.

The Schr\"odinger equations~\eqref{eqn:spatial-schroedinger} are solved by
\begin{equation}
\psi_{\uparrow\downarrow}(t,z)
= \rme^{\rmi \varphi_{\uparrow\downarrow}(t,z)} \chi_{\uparrow\downarrow}(t,z-u^{\uparrow\downarrow}_t) \,,
\end{equation}
where $u^{\uparrow\downarrow}_t$ are the unperturbed classical trajectories solving $m \ddot{u}_t^{\uparrow\downarrow} = \pm F_t$ and the phases are given by
\begin{equation}
\varphi_{\uparrow\downarrow}(t,z)
 = \frac{m}{\hbar} z \dot{u}^{\uparrow\downarrow}_t
 - \frac{m}{2 \hbar} \int_0^t \rmd t' \, \left(\dot{u}^{\uparrow\downarrow}_{t'}\right)^2 \,.
\end{equation}
Note that after a full traversal of the interferometer, when the trajectories are recombined, we have $\varphi_\uparrow = \varphi_\downarrow$ and can ignore this absolute phase contribution.

The wave functions $\chi_{\uparrow\downarrow}$ solve the Schr\"odinger equations in the respective co-moving frames:
\begin{equation}\label{eqn:comoving-schroedinger}
 \rmi \hbar \partial_t \chi_{\uparrow\downarrow}
 = -\frac{\hbar^2}{2m} \partial_z^2 \chi_{\uparrow\downarrow}
 + U^{\uparrow\downarrow} \chi_{\uparrow\downarrow} \,,
\end{equation}
where $U^{\uparrow\downarrow}(t,z) = U(u^\uparrow_t,u^\downarrow_t;z+u^{\uparrow\downarrow}_t)$. In the considered situation of a localized wave function, we can expand the potentials to quadratic order in $z$. Assuming, w.l.o.g., $R \zeta_t := u^\uparrow_t = - u^\downarrow_t > 0$ for the gravity-free trajectories, we find
\begin{subequations}\label{eqn:quadpot}\begin{align}
U^{\uparrow\downarrow}(t,z) &\approx U_0^{\uparrow\downarrow}(t) + U_1^{\uparrow\downarrow}(t) z + U_2^{\uparrow\downarrow}(t) z^2
\intertext{with}
U_0^{\uparrow\downarrow}(t) &= -\frac{G m^2}{R} \left(\frac{6}{5} c_{\uparrow\downarrow} + c_{\downarrow\uparrow} \Xi(\zeta_t)\right) \\
U_1^{\uparrow\downarrow}(t) &= \mp\frac{G m^2}{2R^2} c_{\downarrow\uparrow} \Xi'(\zeta_t) \\
U_2^{\uparrow\downarrow}(t) &= \frac{G m^2}{8R^3} \left(
4 c_{\uparrow\downarrow}
- c_{\downarrow\uparrow} \Xi''(\zeta_t)
\right) \,.
\end{align}\end{subequations}

\section{Gaussian solution}\label{sec:gauss}

We can provide exact solutions of equations~\eqref{eqn:comoving-schroedinger} assuming the quadratic potentials~\eqref{eqn:quadpot} and Gaussian initial conditions. They are given by
\begin{subequations}\begin{equation}\begin{split}
\chi_{\uparrow\downarrow}(t,z)
= (2\pi A^{\uparrow\downarrow}_t)^{-1/4}
\rme^{\rmi \left(\frac{m}{\hbar} \dot{\nu}^{\uparrow\downarrow}_t z + \phi^{\uparrow\downarrow}_t\right)} \\
\times \exp(-\frac{(z-\nu^{\uparrow\downarrow}_t)^2}{4A^{\uparrow\downarrow}_t}\left(1 - \frac{\rmi m}{\hbar} \dot{A}^{\uparrow\downarrow}_t\right)) \,,
\end{split}\end{equation}
with $\nu^{\uparrow\downarrow}_t$ and $A^{\uparrow\downarrow}_t$ solving
\begin{align}
 \ddot{\nu}^{\uparrow\downarrow}_t &= -\frac{1}{m} (U^{\uparrow\downarrow}_1(t) + 2 U^{\uparrow\downarrow}_2(t) \nu^{\uparrow\downarrow}_t) \\
 \ddot{A}^{\uparrow\downarrow}_t &= \frac{\hbar^2 + m^2 \left(\dot{A}^{\uparrow\downarrow}_t\right)^2}{2 m^2 A^{\uparrow\downarrow}_t} - \frac{4}{m} U^{\uparrow\downarrow}_2(t) A^{\uparrow\downarrow}_t
\end{align}
and the time dependent phases given by
\begin{equation}\begin{split}
 \phi^{\uparrow\downarrow}_t = \phi_0 - \frac{m}{2\hbar} \nu^{\uparrow\downarrow}_t \dot{\nu}^{\uparrow\downarrow}_t
\\
-  \int_0^t \rmd t' \left( \frac{\hbar^2}{4 m A^{\uparrow\downarrow}_{t'}}
+ U^{\uparrow\downarrow}_0(t') + \frac{1}{2} U^{\uparrow\downarrow}_1(t') \nu^{\uparrow\downarrow}_{t'}
\right) \,.
\end{split}\end{equation}\end{subequations}
Note that $\chi_{\uparrow\downarrow}(t=0) = \psi_\uparrow(t=0) = \psi_\downarrow(t=0)$ must have the same initial conditions as $\psi_{\uparrow\downarrow}$, implying that the initial phase $\phi_0$ is the same for both spins.

Writing $U_i^{\uparrow\downarrow}(t) =: \hat{U}_i(t) \pm \Delta U_i(t)/2$, $\nu^{\uparrow\downarrow}_t =: \hat{\nu}_t \pm \Delta \nu_t/2$, and $A^{\uparrow\downarrow}_t =: A_f(t) + \hat{A}_t \pm \Delta A_t/2$, with the free spreading $A_f(t) := A_0 + \hbar^2 t^2 / (4 m^2 A_0)$ separated out, and only including terms up to linear order in $G$, we find the phase difference
\begin{equation}\label{eqn:phase-diff}¨
 \Delta \phi_t = \phi^\uparrow_t - \phi^\downarrow_t = \frac{1}{\hbar}\int_0^t \rmd t' \left(
\frac{\hbar^2 \, \Delta A_{t’}}{4 m A_f^2}
- \Delta U_0(t)
\right) \,.
\end{equation}
The equations of motion up to linear order in $G$ are
\begin{subequations}\begin{align}
 \ddot{\hat{\nu}}_t
&= -\frac{\hat{U}_1(t)}{m} \\
 \Delta \ddot{\nu}_t &= -\frac{\Delta U_1(t)}{m} \\
\ddot{\hat{A}}_t &= \frac{\hbar^2}{2 m^2 A_f(t)} \left(1 - \frac{\hat{A}_t}{A_f(t)}\right) - \frac{4}{m} A_f(t) \hat{U}_2(t) \\
\Delta \ddot{A}_t &= \frac{\hbar^2 \Delta A_t}{2 m^2 A_f(t)^2} - \frac{4}{m} A_f(t) \Delta U_2(t)
\end{align}\end{subequations}
with the average potentials and potential differences
\begin{subequations}\begin{align}
\Delta U_0(t) 
&= -\frac{G m^2}{R} \cos(2\alpha) \left(\frac{6}{5} - \Xi(\zeta_t)\right) \\
\hat{U}_1(t) 
&= -\frac{G m^2}{4R^2} \Xi'(\zeta_t) \\
\Delta U_1(t) 
&= \frac{G m^2}{2R^2} \cos(2\alpha) \Xi'(\zeta_t) \\
\hat{U}_2(t) 
&= -\frac{G m^2}{16R^3} \left(4 - \Xi''(\zeta_t)\right) \\
\Delta U_2(t) 
&= -\frac{G m^2}{8R^3} \cos(2\alpha) \left(4 + \Xi''(\zeta_t)\right) 
\,.
\end{align}\end{subequations}

As a further approximation, we assume that the initial and final acceleration stages with $F_t \not= 0$ in order to separate the trajectories are very short compared to a free flight stage in between. We can then neglect the gravitational effects during the acceleration stages and evaluate the potentials for two trajectories at the distance $d \geq 2R$. This implies $\zeta_t = d/(2R) \geq 1$ and thus the constant potentials
\begin{subequations}\label{eqn:const-potentials}\begin{align}
\Delta U_0
&\approx -G m^2 \left(\frac{6}{5R} - \frac{1}{d}\right) \cos(2\alpha)  \\
\Delta U_2
&\approx -G m^2 \left(\frac{1}{2R^3} + \frac{1}{d^3}\right) \cos(2\alpha) \,.
\end{align}\end{subequations}
For experiments with flight time $\tau \ll 1/\omega := 2\sqrt{2} m A_0/\hbar$, such that the wave function spreading can be neglected, i.e.\ $A_f(t) \approx A_0$, we then find the solution
\begin{equation}
 \Delta A_t \approx -\frac{\hbar \Delta U_2}{\sqrt{2} m^2 \omega^3} \sinh^2(\omega t) \,.
\end{equation}
The phase integral~\eqref{eqn:phase-diff} becomes approximately
\begin{equation}
 \Delta\phi_t \approx -\frac{t}{\hbar} \Delta U_0
 + \frac{2\omega t - \sinh(2\omega t)}{2\sqrt{2} m \omega^2} \Delta U_2 \,,
\end{equation}
which after expanding up to third order in time and inserting the potentials~\eqref{eqn:const-potentials} results in the phase difference
\begin{equation}\label{eqn:phase-diff-final}
 \Delta\phi_t \approx \frac{G m^2 t}{\hbar} \cos(2\alpha) \left(
\frac{6}{5 R}
- \frac{1}{d}
+ \hbar^2 t^2 \frac{\tfrac{1}{R^3} + \tfrac{2}{d^3}}{12 m^2 A_0}
\right) \,.
\end{equation}
In the regime where $d \gg R$, the middle term is negligible. The ratio of the third to the first term is of the order of $(A_f - A_0)/R^2$ in this regime and, hence, also negligible as long as the assumption of a narrow wave function compared to the particle radius remains valid.

Finally, for the sake of providing numerical estimates, we can express the mass in terms of the particle's radius and density, leaving us with
\begin{align}
 \Delta \phi_t &\approx \frac{32 \pi^2}{15 \hbar} G \rho^2 R^5 t \cos(2\alpha) \nnl
&\sim \unit{10^2}{\power{\micro\meter}{-5}\power{\second}{-1}} \times R^5 t \cos(2\alpha)
\end{align}
for typical densities around $\unit{4 \pm 2}{\gram\per\centi\meter\cubed}$.

\section{Discussion}\label{sec:discussion}

With equation~\eqref{eqn:phase-diff-final} we have derived a compact result for the self-gravitationally induced phase difference in the experimentally important situation of a superposition of two narrowly peaked wave packets along separate trajectories. The considered approximations are valid as long as these wave packets remain much smaller than the particle radius and the trajectories are separated by some $d \geq 2R$ in a short time compared to the entire flight time through the interferometer.

This result agrees~\footnote{Note that there is a mistake in~\cite{grossardtDephasingInhibitionSpin2021} that omits the relevant $\Delta U_0$-proportional term in the phase integral~\eqref{eqn:phase-diff} which is missing from eq. (30) of~\cite{grossardtDephasingInhibitionSpin2021}.} with the earlier calculations~\cite{hatifiRevealingSelfgravitySternGerlach2020}. In reference~\cite{aguiarProbingGravitationalSelfdecoherence2023}, the self-energy contribution to the potential has been neglected without justification, resulting in a phase difference according to only the mutual energy, the second term in equation~\eqref{eqn:phase-diff-final}. It is argued there, that this would provide a lower limit on the observed phase difference. However, as is obvious from equation~\eqref{eqn:phase-diff-final}, the self-energy and mutual energy enter the phase difference with opposite sign and could in principle even cancel out in an experiment where $5 R = 6 d$. In order to maximize the observed phase difference, the distance $d$ between the trajectories should be as large as possible in order to render the mutual energy contribution negligible.

Finally, the gravitational self-energy is known to be increased if accounting for the atomic substructure of the material for a sufficiently well localized (below the Debye-Waller length scale) wave function~\cite{yangMacroscopicQuantumMechanics2013,grossardtEffectsNewtonianGravitational2016,grossardtOptomechanicalTestSchrodingerNewton2016}. This yields an additional contribution to the phase difference from the atomic self energies:
\begin{equation}
 \Delta \phi_\text{atom} \approx \sqrt{\frac{2}{\pi}}\frac{G m m_\text{atom} t}{\hbar \sigma} \cos(2\alpha) \,,
\end{equation}
with $m_\text{atom}$ the mass of the atomic constituents and $\sigma = 2 \pi \sqrt{B}$ with $B$ the Debye-Waller factor. For realistic parameters, however, $\Delta \phi_\text{atom}$ only yields a relevant contribution for a particle radius $R \lesssim \unit{1}{\nano\meter}$. Above that size, the phase difference is dominated by the first term in equation~\eqref{eqn:phase-diff-final}.

As expected, the phase difference disappears for a symmetric superposition $\alpha = \pi/4$ and is maximal if one of the coefficients $c_{\uparrow\downarrow}$ becomes as small as experimentally possible while still allowing for sufficient statistics.

\end{document}